# AN ALGORITHM FOR OPTIMIZED SEARCHING USING NON-OVERLAPPING ITERATIVE NEIGHBOR INTERVALS

Elahe Moghimi Hanjani and Mahdi Javanmard

Department of computer engineering, Payam Noor University, Tehran, Iran,
Po Box 19395-3697
elahe_moghimi@yahoo.com
info@javanmard.com

## ABSTRACT

*We have attempted in this paper to reduce the number of checked condition through saving frequency of the tandem replicated words, and also using non-overlapping iterative neighbor intervals on plane sweep algorithm. The essential idea of non-overlapping iterative neighbor search in a document lies in focusing the search not on the full space of solutions but on a smaller subspace considering non-overlapping intervals defined by the solutions. Subspace is defined by the range near the specified minimum keyword. We repeatedly pick a range up and flip the unsatisfied keywords, so the relevant ranges are detected. The proposed method tries to improve the plane sweep algorithm by efficiently calculating the minimal group of words and enumerating intervals in a document which contain the minimum frequency keyword. It decreases the number of comparison and creates the best state of optimized search algorithm especially in a high volume of data. Efficiency and reliability are also increased compared to the previous modes of the technical approach.*

## 1. INTRODUCTION

The most exceptional search engine would not provide good quality results if the original keywords selected by the user were not suitable. Therefore, the proposed algorithm aims at searching on a set of alternative keywords generated based on a user's original keywords to help a user in his/her subsequent search activities and reduce the time with using the relative range in searching.

Plane sweep algorithm considers that keywords which appear in the neighbourhood in a document that are related. Therefore we use position of keywords in a document as the unit of queries. By considering keyword positions we can find a paragraph or a sentence in a document which describes what we want to know. We define ranks of regions in documents which contain all specified keywords in order of their sizes. This is called proximity search [4,5,8]. Proximity searching deals with the location and the distance relationship among keywords in a document.

The algorithm can find a target item in an offsetList that is selected according to the minimum keyword and distance factor, one can trade accuracy for speed and find a part of the offsetList containing the target item even faster. Moving iteratively builds a sequence of solutions generated by the algorithm. It requires finding all the partial range for a given offset list using the technique to find the keyword with minimum repeat.

We implemented and tested a unified approach to generating query suggestions based on phrases. We perform a search for all subsets at query time, in a case which a word overlaps itself repeatedly, we count the number of ordered pairs of symbols that are adjacent in the document and by using the iterated partial Search, and we limit the search space. It will be performed in less time especially at high data storage.

Running time of the proposed algorithm can be achieved in time $\mathrm{o}((n-\alpha)\log k)$, where n is the frequency of keywords occurrence in a document, $\alpha$ is the frequency of tandem replicated data and k is the number of query terms in a query.

## 2. RELATED WORKS

In string matching, there are some results on finding k keywords within a given maximum distance d. Gonnet etal. proposed an algorithm for finding two keywords P1 and P2 within distance d in $\mathrm{o}(m_1 + m_2)$ time, where $m_1 < m_2$ are the numbers of occurrences of the two keywords. Baeza-Yates and Cunto proposed the abstract data type Proximity and an $\mathrm{o}(\log n)$ time algorithm. Manber and Baeza-Yates also proposed an $\mathrm{o}(n \log n)$ time algorithm [4,5].

They assume that the maximum distance d is known in advance. Sadakane and Imai proposed an $\mathrm{o}(n \log k)$ time algorithm for a restricted version of the problem. Plane sweep algorithm achieves the same time complexity for the k-keyword proximity problem while dealing with a generalized version of the problem [4,5].

There are several algorithms for finding exact tandem repeats. Most of these algorithms have two phases, a scanning phase that locates candidate tandem repeats, and an analysis phase that checks the candidate tandem repeats found during the scanning phase [10]. We use the concept of finding repeated data in pre-processing phase to decrease the running time of plane sweep algorithm. In partial search of our algorithm, one is interested in finding the exact address of the target item and search around it. So, the proposed algorithm limits the searched area to a subset of document with escaping from some ineffective keywords, which reduce running time in plane sweep algorithm.

## 3. PROPOSED ALGORITHM

Keyword proximity searching in a document is the method to find the relevant document that all the terms in a query appear with in a relatively small fixed-size window.

The notion of proximity search differs from text search with wildcards in three key ways:

1. The total length of the matched string is bounded, thus there is a cumulative bound on the length of the arbitrary sequences.
2. The order of the search terms is not specified; thus, in the example, any permutation of A, B, and C is permitted.
3. Search begins with an index (i.e., a list of occurrences for each term, often called a position list) instead of the original text.[4]

Plane sweep algorithm search the inverted lists until the search range is detected, but the proposed algorithm search in the neighbourhood of the minimum keyword then after scanning the search range and getting the critical range which is minimal, the algorithm shifts to the next range. The important point is that we remove the sequences which contain the replicated data and also allow related ranges which is presented according to the algorithm and have significantly reduced the expected conditions to the previous algorithms according to the above parameters.

We try to have a relationship based on query and keywords that we find in the search document, and also we need to reduce the number of comparisons so that search operation performs faster. The proposed algorithm reduces the searched area to a minimum and relies on an optimized search algorithm for effectively pruning the search space.

**Definition1**. Given $k$ keywords $W_1, W_2, ..., W_k$, a set of lists $K = \{K_1, K_2, ..., K_k\}$ where the i-th keyword $W_i$, and positive integers $f_1, f_2, ..., f_k$, and $k (\leq k')$, the generalized $k$ keyword proximity problem is to find the smallest range that contains k _ distinct keywords $W_i (1 \leq i \leq k)$ appear at least $f_i$ times each in the range.

Note that the problem becomes the basic plane sweep algorithm when all $f_i = 1$ for $1 \leq i \leq k$. If we show each offset list with w and frequency as f, we have, $w[i] = w[i + f]$ ,$1 \leq i \leq |w| - \alpha$. The offset of w for replicated tandem word is the offset of the first word location.

**Definition2**. Let $(X, d)$ be a metric space and $R^* \in D$ the set of offsets, a range query $ffset[Min(Q)] = q$ , $(q, r), q \in D, r \in R^+$ , reports all keywords that are within distance $r$ to $q$, that is $(q, r) = \{u \in R^*, d(u, q) \leq r\}$ . The volume defined by $(q, r)$ is the range space, and all the keywords from $R^*$ are reported. The proposed algorithm can be implemented using range queries.

**Definition3**. A range space is a set $\Sigma = (D, R^*)$ , where D is the search space and $R^*$ is the family of subset of D. The elements of $R^*$ are ranges of $\Sigma$ , where $\Sigma$ is a finite range. In optimization queries, we want to return an object that satisfies certain condition with respect to the query range. Ineffective searches have no effect on the result. So we have:

$$F(D) = \bigcup_{R^* \in D} (R^*) , \ F(D) = \{X \in D \mid X = R^*_{j1}, R^*_{j2}, ..., R^*_{jk}\}$$

D is the search space which contains ranges that can be matched with chains of basic moves.

**Definition4**. Let $R_D$ be the precision of the ranges that are relevant to the query in the algorithm and also retrieved from the plane sweep algorithm, in which a better solution has been found. $R_D$ in a document can be defined as the conditional probability which denotes the probability that ranges have within a document. This parameter shows the ratio of relevant ranges.

$$R_D = \frac{\#(Retreived.ranges)}{\#(Plane - sweep.Candidate.ranges)} = \qquad\qquad (1)$$

$$\frac{|Relevent.ranges \cap Plane - sweep.Candidate.ranges|}{|Plane - sweep.Candidate.ranges|}$$

First, we have tried to find replicated words in a list of tandem words with the specified offset which is the output stage of the preprocessing and then we consider the keywords which have minimum counter and limit our search area to the range around that keyword. We reduce the number of comparisons with counting the number of tandem replicated words and also removing unsatisfied searches. The main objective is to find the most efficient and relevant answer for the query.

There are so many results which contain the query's keywords but users are interested in a much smaller subset. For this purpose we define a range as $R^*$ where it contains nearest neighbours of minimum counted keyword, given the proposed algorithm, we can perform a local search in $R^*$, as we should consider the range $Pos[M_i] + (|Q| - 1)$, $Pos[M_i] - (|Q| - 1)$ around the minimum keyword.

We want to explore $R^*$ using a walk that steps from one $R^*$ to a "nearby" one, with the list of defined minimum keyword. This algorithm escapes from ineffective keywords by applying the current partial range.

If we show distance factor with $DF_{Minkeyword}$ which is the distance between two minimum keyword and defined as $DF_{Minkeyword} = \lfloor DF_{K_{1,2}}, DF_{K_{2,3}}, ...., DF_{K_{n,m}} \rfloor$, where m=#ofMinKeyword, according to our condition it might be greater than $|Q| - 1$, where $|Q|$ is the length of query, and if one of the distance factor is lower than $|Q| - 1$, we should escape the right range of $K_{M_{i-1}}$ or left range of $K_{M_i}$ to have the minimum comparison, in this situation the search range would not overlap.

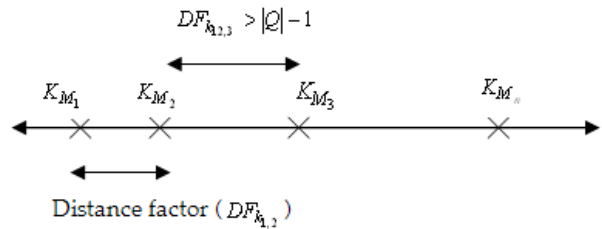

Fig.1 Distance factor between two minimum keywords.

It is possible that there are few keywords in the document with the minimum number. We consider another factor as distance factor. The distance factor is the number of locations between the two minimum keywords, the value must be greater than the Query length ($|Q| - 1$) otherwise the search range may overlap And search results will not be optimal, if this factor is also the same, we consider one of the keywords with minimum number randomly and the search range limited to the distance around that keyword.

The implementation of this algorithm is intended to count the number of tandem repeat words in a document using inverted file and also limit the search range using the technique of partial range.
In this algorithm, two pointers are used to search a document that Scan offset list from left to right, Pointer $P_l$, which refers to the left of the offset and $P_r$ which is refer to the last offset in the range.

As we defined $R^*$ as a searched range, Iterated partial search achieved the results as follows:
Set $P_l, P_r$ with the position of the first $R^*$, then iteratively moves to the next $R^*$ that contains the item of interest. It means the ranges that are irrelevant for the result is removed.

$P_r$ moves in a defined range, if its value is greater than the range size, it returns to the next beginning of the interval which $P_l$ is pointed. After a critical range was set, $P_l$ also move forward one place at the offset list. In this algorithm, counter is stored for each offset value. And if its value is greater than one, compare operations do not need to do and we only move forward according to the size of the defined range, the algorithm doesn't consider all position of word in a document. In fact, we skip the repetitive sequences. After each critical range is defined, minimality is also checked. This continues until we reach end of the list.

Our work tries to improve the plane sweep algorithm by efficiently calculating the minimal group of words as a result.

The most important aspect to measure, apart from the accuracy, would be the time required for the algorithm to compute the recommendations. Apart from that we also need to measure the number of iterations required by users to reach the desired results. This can be compared with the number of iteration required by the user in an environment without removing ineffective search and also with replicated data. In practice, there is a trade-off between the encoding overhead and the amount of achievable reduction.

Number of comparison in a defined ranges and also number of comparison in replicated word is defined below:

$$\beta = \sum_{k=1}^{|\#ofMinKeyword|} \left( \sum_{j=1}^{|D|} \left( \sum_{i=1}^{|Q|} (i \times \pi_i) \right) \right), \gamma = \sum_{j=1}^{|\alpha|} \left( \sum_{i=1}^{|Q|} (i \times \pi_i) \right)$$

The number of comparison for replicated word is calculated from following equation:

$$\alpha = \sum_{k=1}^{|D|} (ctr_k - 1), \text{ if k>1.} \tag{2}$$

$\alpha$ = Number of Replicated word
|D| = Length of Document offset list - |Q| -1
|Q| = Length of Query

$$\pi_i = \begin{cases} 1 & \text{if match occurred in Doc list} \\ 0 & \text{O.W} \end{cases}$$

Let $\pi_i$ denote the Availability factor of a word in a query, which is describing the number of comparisons made. $\pi_i$ set to one if position i considered with the algorithm, otherwise, the number of comparison is zero and it is also set to zero, so only position i is applied in Formula(3).

As a consequence of the algorithm and basic plane sweep and number of tandem replicated word, and also removing the comparison of ineffective keywords, number of comparisons related to the algorithm can be formulated as follows:

$$C_n = \beta - \gamma \tag{3}$$

As shown in formula (3) number of comparison in the proposed algorithm is the sum of ranges which in each range number of comparison is the difference between the number of comparison in plane sweep algorithm and the number of comparison in replicated words. So according to formula (3), we reduce the number of comparison in our proposed algorithm.

For example, consider the following document:

Document :{ ABCCCABCCBACBBBCBA}

| Keyword in Doc | A | B | C | A | B | C | B | A | C | B | C | B | A |
|---|---|---|---|---|---|---|---|---|---|---|---|---|---|
| # of repeated words | 1 | 1 | 3 | 1 | 1 | 2 | 1 | 1 | 1 | 3 | 1 | 1 | 1 |

Fig. 2 searching on document { ABCCCABCCBACBBBCBA}

In this example number of A is 4, number of B is 5, and number of C is 4. The minimum number of tandem repeated word is belonging to A and C.

We consider the distance factor as distance between two same keywords. The distance between each A is $D_A = [2,3,4]$ and the distance between each C is shown in vector $D_C = [1,2,1]$.

In the second case, the distance between $D_{C_{1,2}}$ and $D_{C_{3,4}}$ is lower than the query length ($|\varrho|-1$) and removed from the search so, search will continue around A. Ranges specified in fig 3, which is shown as $I_1, I_2, I_3, I_4$.

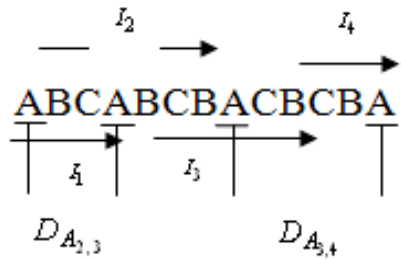

Fig. 3 Example of searching with considering minimum repeat words

In this section we describe a proposed algorithm and show its improved average running time. The algorithm is defined below:

(1) Sort offset of keyword $P_{ij} = (j = 1...n)$ in a document in increasing order, we also add counter for replicated tandem words to the list.
(2) Add the number of each query's keywords in a offsetList
(3) Get the minimum replicated keyword from the list considering the distance factor
(4) Repeatedly increase i by one until we get end of the Min-keyword offsetList
(5) If we have passed end of the list, sort interval in a heap with considering the tandem replicated word counter and output them, finish.
(6) Considering the neighbor of the target minimum keyword according to distance factor so that we have no overlap range.
(7) Repeatedly increase $P_i$ by one until the current range is a candidate range or we have passed the end of the list.
(8) If we have passed end of the partial list, Go to step 4.

(9) Repeatedly increase $P_r$ by one until the current range is not a candidate range.

(10) The range $(P_l, P_{r-1})$ is a critical range. Compare the size of range $(P_l, P_{r-1})$ with the stored minimum range and replace the minimum range with the current range if $(P_l, P_{r-1})$ is smaller.

(11) Go to step 4.

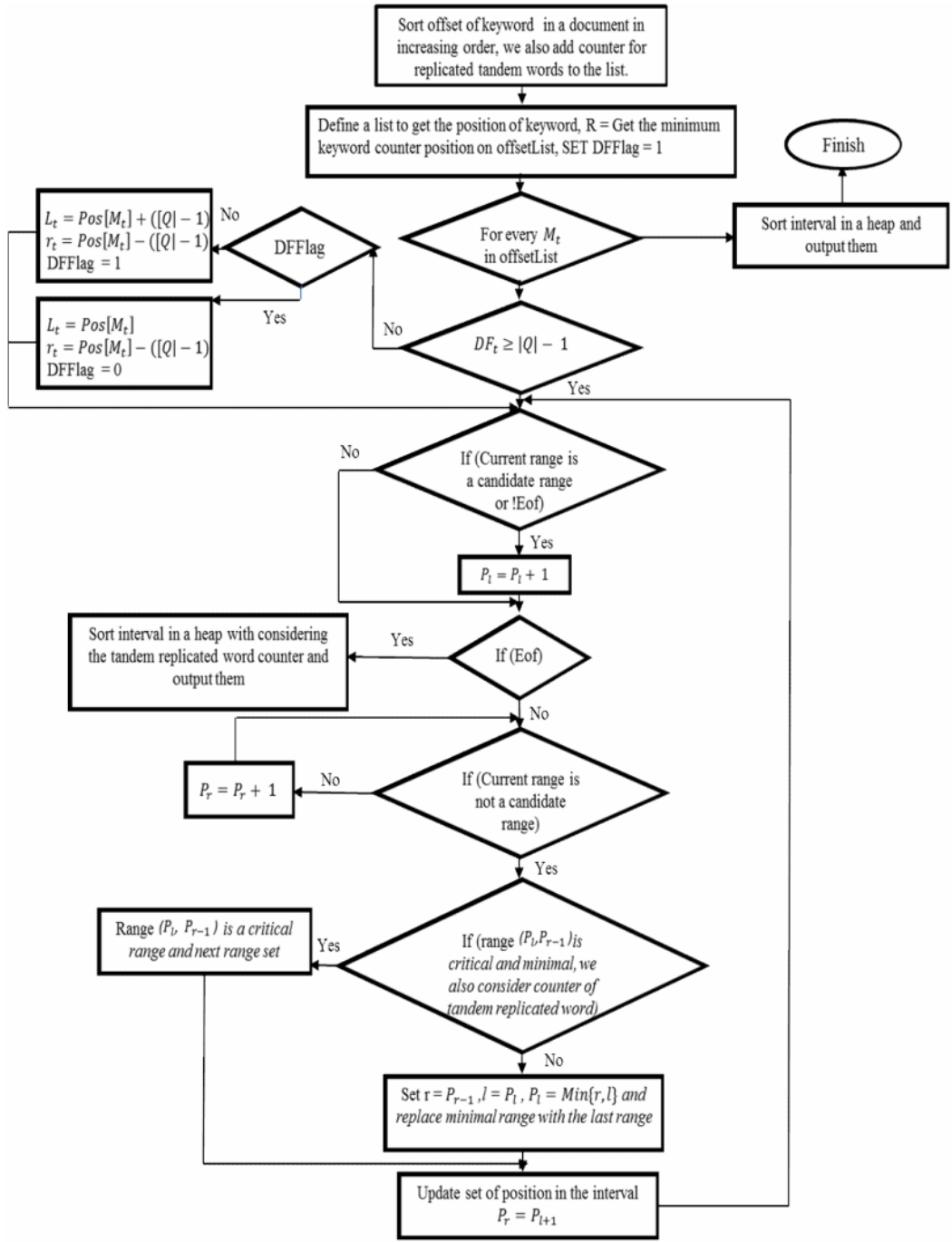

Fig. 4 Proposed algorithm flowchart

Suppose we have a query of "BCA" that is searched in the following document offset list, the algorithm needs to compute all the related keywords:

Document :{ CABABABCABBB}

A 'tandem replicated word' is a string of the form

$X^{s_\alpha} Z^{y_\alpha} ... X^{s_\alpha} W^{q_\alpha}$

Where $s_\alpha, y_\alpha, q_\alpha \geq 1$ for some $\alpha \in \{1,2,...,n\}$.

The concatenations of the tandem replicated word is as follows:

$\prod_{i=1}^{13} W^{q_i} X^{r_i} Z^{y_i} X^{r_i} Z^{y_i} X^{r_i} Z^{y_i} W^{q_i} X^{r_i} Z^{y_i}$

Where $q_i \in \{1\}$; $r_i \in \{1\}$; $y_i \in \{1,3\}$.

The frequency of replicated words in the list is: $W_{sim} = \dfrac{\sum |\alpha|}{|D|} = \dfrac{2}{14} = 0.14$ .

The ratio of above example is calculated which is based on the solution ranges in a document in the proposed algorithm and plane sweep algorithm. The less ratio value is shown the greater efficiency of the proposed algorithm because it reduces the number of comparisons and as a result the number of ranges.

Any offset in the following list are:

$K_1 \in \{2,4,6,9\}$, $K_2 \in \{0,7\}$ and $K_3 \in \{1,3,5,8\}$.

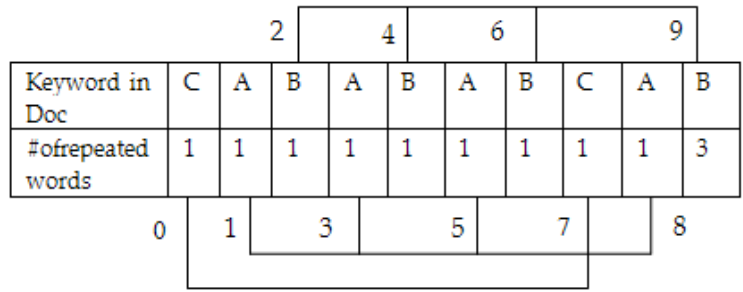

Fig.5 Searching ranges on $K_1, K_2, K_3$

Search has been defined from left to right with respect to the range for 3-keyword. There are two partial range in the offset list, $R_1^*$ and $R_2^*$.

Where :

$R_1^* = CAB, R_2^* = ABCAB$

Finding the solution for the above example:

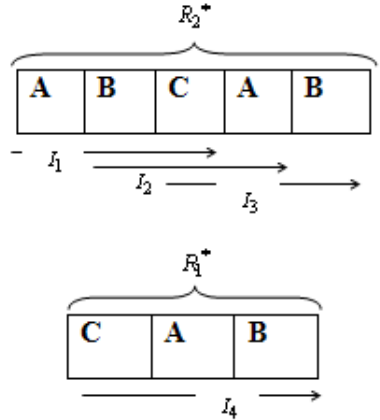

Fig.6 Result of the algorithm is shown as $I_1, I_2, I_3, I_4$

$R_1^*, R_2^*$ are the partial ranges that are the output of the algorithm and $I_1, I_2, I_3, I_4$ are the result range.

**Definition5**. Let $W_{sim}$ be the repetition factor of the tandem replicated word. It is defined below:

$$W_{sim} = \frac{\sum |\alpha|}{|D|}, 0 \leq W_{sim} \leq 1 . \tag{4}$$

Where $|D|$, is the size of the offset list, and $|\alpha|$ is the frequency of replicated words in the list.
This parameter shows the ratio of similarity that was retrieved from the keywords in a document.

## 4. TEST RESULTS

In order to study the effect of tandem replicated word with iterative partial range, we ran some experiment on a different lists size, and the result is shown below. The data sets and test results are used to assess performance measures for the algorithm under test.

Experiments (Table1) views the sequence as it has been produced by a random file with the different repetition factor of tandem replicated word on different file size. It must be noted that the run time depends not only on repetition factor of the input document, but also on the number of keywords in query. In fact, if repetition factor (it means the number of tandem repeat words in the document), and also the number of keywords in question is greater, the proposed algorithm work better.

Table1: Results of the tests on 3-keywords

| $W_{sim}$ / Data size | 500 | 1000 | 2000 | 3000 | 4000 |
|---|---|---|---|---|---|
| 0.2 | 0.0135 | 0.0218 | 0.0337 | 0.0443 | 0.0535 |
| 0.4 | 0.0163 | 0.0234 | 0.0352 | 0.0474 | 0.0590 |
| 0.6 | 0.0171 | 0.0244 | 0.0381 | 0.0527 | 0.0668 |

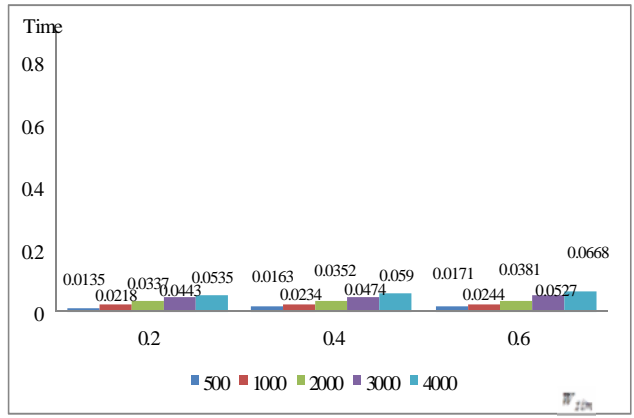

Fig. 7 Number of Comparisons on 3-keywords query.

We compared the three search algorithm, plane sweep, WPSR and modified WPSR on the data set and the result is shown on Table2.

Table2: Results of the tests on 3 algorithms with 3-keywords query and, $R_D = 0.6$, $w_{sim} = 0.6$.

| Datasize | Plane sweep | WPSR | M-WPSR |
|----------|-------------|--------|--------|
| 3000 | 0.1495 | 0.0790 | 0.0785 |
| 6000 | 0.1654 | 0.0809 | 0.0800 |
| 12000 | 0.3114 | 0.1500 | 0.1490 |
| 18000 | 0.4764 | 0.2282 | 0.2197 |
| 24000 | 0.594 | 0.2819 | 0.2801 |
| 30000 | 0.7161 | 0.3548 | 0.3448 |

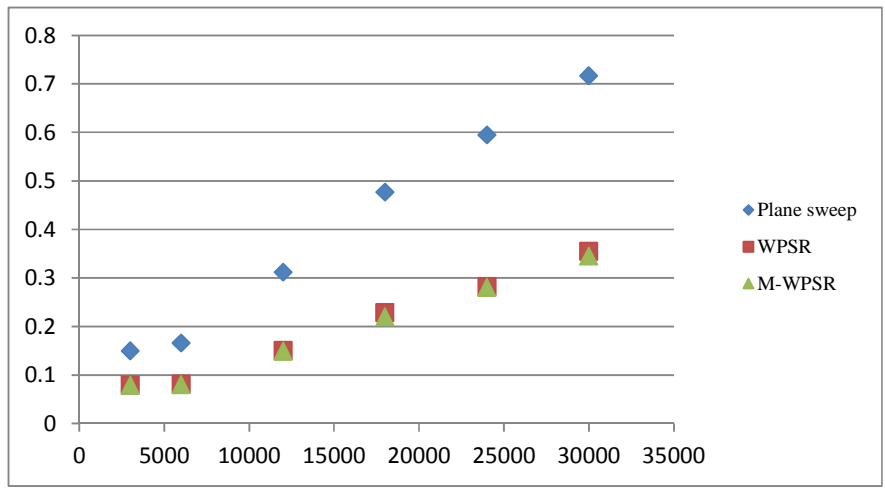

Fig. 8 Number of Comparisons made by 3 algorithms with
3-keywords query and query and, $R_D = 0.6$, $w_{sim} = 0.6$.

Table3: Results of the tests on offsets

| Datasize | Plane sweep | M-WPSR |
|----------|-------------|--------|
| 900 | 0.0239 | 0.0216 |
| 1800 | 0.0424 | 0.0350 |
| 3600 | 0.0775 | 0.0600 |
| 7200 | 0.1449 | 0.1101 |
| 14400 | 0.2751 | 0.2110 |
| 28800 | 0.5584 | 0.4202 |

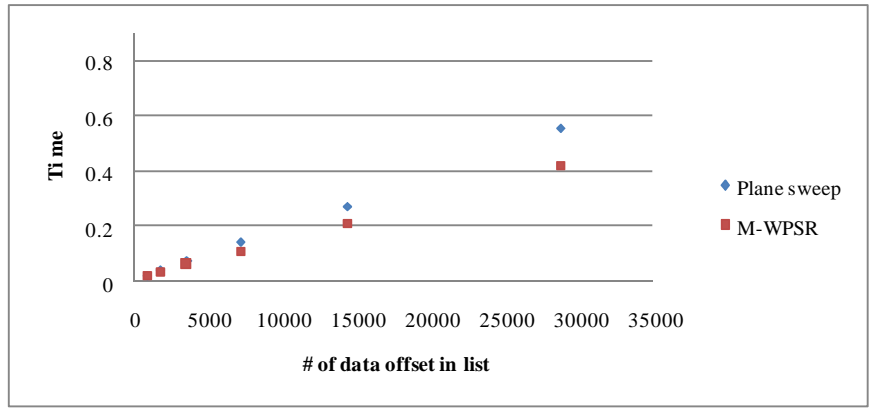

Fig. 9 Number of Comparisons made by modified WPSR and plane sweep
algorithm with 3-keywords query and $W_{sim} = 0.4$ .

The experiment is shown above for the three search algorithm on the different data sets and with the different parameter of $W_{sim}$ and $R_D$ , and also with changing the number of keywords in a query, we see that it leads to a better running time especially in high volume repetitive data.

The proposed algorithm is expected to improve search accuracy and effectiveness, with the fast-growing availability of information online, who users may not be aware of most the updated critical keywords, the proposed system is also expected to improve search efficiency. Finally, we observe that using tandem replicated word with iterative partial range make all size of the random sample better, especially in a large size, and it leads to a better running time.

## 5. CONCLUSION

We remove the sequences which contain the replicated data and also allow related ranges which are presented according to the proposed algorithm. This has significantly reduced the expected conditions to the previous algorithms according to the above parameters.

The proposed algorithm will be more optimized when the number of data is increased, because in this situation, number of retrieved ranges is decreased more by mentioned algorithm comparisons. The effect of algorithm on high volume of data is more significant, and it is robust, and highly effective. Experimental results show that the new algorithm performs well in practice.

## AUTHORS

**Elahe moghimi hanjani** received her B.Sc. in computer engineering from Azad University central Tehran branch, Tehran, IRAN, in 2008, and is a M.Sc. student in computer engineering in Payam Noor University (PNU). Her research interests include optimizing text retrieval algorithm, data mining.

**Mahdi javanmard** received his M.Sc. degree in Electrical Engineering from the University of New Brunswick, Canada and Ph.D. degree in Electrical and Computer Engineering from Queen's University at Kingston, Canada, in 2002 and 2007 respectively. He is a faculty member of Payam Noor University (PNU) and currently Head of COMSTECH Inter Islamic Network on Virtual Universities (CINVU). He has been teaching for many years at different universities where he has been involved in their course development for the Computer Science Department. Additionally, he works as a System Development Consultant for various companies. Dr Javanmard's research interests are in the areas of Information & Communication Security, Speech Recognition, Signal Processing, Urban Management & ICT, and Ultrasound Medical Imaging.